\documentclass[                     
               showpacs,            
               preprintnumbers,     
               aps,                 
               prd,                 
               nofootinbib,         
               floats,floatfix      
               ]{revtex4}

\usepackage{amssymb,graphicx}

\newcommand{\im}{\mbox{Im}}

\newcommand{\Real}{\mathbb{R}}

\begin{document}

\title{Improved outer boundary conditions for Einstein's field equations}

\date{\today}
\author{Luisa T. Buchman$^1$ and Olivier C. A. Sarbach$^2$}
\affiliation{$^1$Center for Relativity, University of Texas at Austin,
1 University Station C1606, Austin, Texas, 78712-1081, USA}
\affiliation{$^2$Instituto de F\a'{\i}sica y Matem\a'aticas,
Universidad Michoacana de San Nicol\a'as de Hidalgo, Edificio C-3, 
Cd. Universitaria. C. P. 58040 Morelia, Michoac\a'an, M\a'exico}

\begin{abstract}
In a recent article, we constructed a hierarchy ${\cal B}_L$ of outer
boundary conditions for Einstein's field equations with the property
that, for a spherical outer boundary, it is perfectly absorbing for
linearized gravitational radiation up to a given angular momentum
number $L$. In this article, we generalize ${\cal B}_2$ so that it can
be applied to fairly general foliations of spacetime by space-like
hypersurfaces and general outer boundary shapes and further, we
improve ${\cal B}_2$ in two steps: (i) we give a local boundary
condition ${\cal C}_2$ which is perfectly absorbing {\em including}
first order contributions in $2M/R$ of curvature corrections for
quadrupolar waves (where $M$ is the mass of the spacetime and $R$ is a
typical radius of the outer boundary) and which significantly reduces
spurious reflections due to backscatter, and (ii) we give a non-local
boundary condition ${\cal D}_2$ which is exact when first order
corrections in $2M/R$ for both curvature {\em and} backscatter are
considered, for quadrupolar radiation.
\end{abstract}


\maketitle

\section{Introduction and main results}
\label{Sect:Intro}

Most often, numerical simulations of black hole spacetimes involve
replacing the physical unbounded domain with a compact domain $\Sigma$
with artificial outer boundary $\partial\Sigma$. In order to obtain a
well-defined Cauchy evolution, boundary conditions need to be
specified at $\partial\Sigma$. Ideally, these conditions should be
formulated in such a way that the artificial boundary is completely
transparent to the physical problem on the unbounded domain. However,
in reality, one aims for boundary conditions which (i) form a
well-posed initial boundary value problem (IBVP) and (ii) insure that
very little spurious reflection of gravitational radiation occurs from
the outer boundary. Boundaries which meet these two criteria are
called absorbing.

The formulation of absorbing outer boundary conditions in general
relativity has always been a challenging problem, primarily because
constraint violations can enter the numerical domain via the outer
boundary, the definition of in- and outgoing radiation is ambiguous,
and the evolution of the geometry of the outer boundary is not known
{\em a priori}. Strides towards the development of absorbing outer
boundaries which both preserve the constraints and control the
gravitational radiation by freezing the Newman-Penrose Weyl scalar
$\Psi_0$ to its initial value have been made in Refs.
\cite{hFgN99,lKlLmSlBhP05,oSmT05,lLmSlKrOoR06,oR06,gNoS06}. For other
approaches which preserve the constraints exactly but control the
radiation at the boundary through different means, see
Refs. \cite{mBbSjW06,hKjW06,mBhKjW07,bZePpDmT06}. In \cite{jNsB04},
boundary conditions which are perfectly absorbing for quadrupolar
solutions of the flat wave equation are analyzed and numerically
implemented via spectral methods, and proposed to be used in a
constrained evolution scheme of Einstein's field equations
\cite{sBeGpG04}. Other methods avoid introducing an artificial outer
boundary altogether by compactifying spatial infinity
\cite{mClLiOrPfPhV03,fP05} or by making use of hyperboloidal slices
and compactifying null infinity (see, for instance
\cite{hF81,jF98,sHcStVaZ05}).

Recently, a hierarchy ${\cal B}_L$ of local boundary conditions for
$\Psi_0$ has been proposed (see Eq. (128) of \cite{lBoS06}) which, for
$L\ge2$, improves the freezing $\Psi_0$ boundary condition given in
\cite{hFgN99,lKlLmSlBhP05,oSmT05,lLmSlKrOoR06,oR06,gNoS06} in that it
is perfectly absorbing on a spherical boundary for linearized
gravitational perturbations with angular momentum number $\ell \le
L$. If the outer boundary is placed in the wave zone (meaning $kR \gg
1$, where $R$ is a characteristic radius of the outer boundary and $k$
is the characteristic wave number) and the solution is smooth enough,
the few lower $\ell$ multipoles will dominate. Therefore, the boundary
condition ${\cal B}_L$ with $L=2,3$ or $4$ should yield very little
spurious reflections in linearized gravity. When considering the full
nonlinear problem, if the outer boundary is not only in the wave zone
but also far away from the strong field region (meaning $R \gg M$
where $M$ is the total mass of the system, that is, the
Arnowitt-Deser-Misner (ADM) mass), one expects the spacetime to be
described by a linear perturbation about Minkowski
spacetime. Therefore, in the nonlinear case, the boundary conditions
${\cal B}_L$ for $L=2,3$ or $4$ should also yield small spurious
reflections provided that $kR \gg 1$ and $R \gg M$.

In this article, we formulate ${\cal B}_L$ to be applicable to more
general slicings than in \cite{lBoS06}; furthermore, we construct
${\cal B}_L$ in such a way that the outer boundary is not restricted
to be an approximate metric sphere. The approach we adopt is to
describe a general four dimensional metric near the outer boundary of
the computational domain as the Schwarzschild metric plus a
perturbation. We perturb Schwarzschild, rather than Minkowski,
spacetime in order to incorporate leading order effects in $M/R$ into
our boundary conditions. For an asymptotically flat spacetime, the
$M/R$ correction terms to Minkowski spacetime are given by the
monopole contribution of the Schwarzschild metric. When these terms
are taken into account, a gravitational wave solution on Minkowski
spacetime acquires two types of correction terms. The first is the
curvature correction term which obeys Hugyens' principle and the
second is a fast decaying term which violates Hugyens' principle and
describes the backscatter off of curvature.

A method which provides exact outer boundary conditions for linearized
waves propagating on a Schwarzschild background has been recently
presented in \cite{sL04a,sL04b,sL05}. However, in view of the
nonlinear theory, there is no advantage to obtaining a boundary
condition which takes into account the exact form of the Schwarzschild
metric beyond the order of $M/R$, since the full spacetime agrees with
the Schwarzschild metric only up to order $M/R$. If second order
effects are to be taken into account, then linear effects in $J/R^2$
(where $J$ is the total angular momentum of the system) and quadratic
effects in $M/R$ from the full metric should also be included.

Our boundary conditions ${\cal B}_L$ are formulated as follows. We
assume that in the vicinity of the outer boundary, the spacetime
metric is the Schwarzschild metric of mass $M$ plus a small
perturbation of it. More precisely, we assume that the spacetime
manifold near the outer boundary can be written locally as a product
of a two-dimensional manifold $\tilde{\cal M}$ and a unit two-sphere
$S^2$ such that if $(x^a)$ denote local coordinates on $\tilde{M}$ and
$(x^A) = (\vartheta,\varphi)$ denote the usual angular coordinates on
$S^2$, the components of the full metric $g^{(full)}$ satisfy the
following conditions:
\begin{equation}
| g^{(full)}_{ab} - \tilde{g}_{ab} | \ll 1, \qquad
| g^{(full)}_{aB} | \ll 1, \qquad
| g^{(full)}_{AB}/r^2 - \hat{g}_{AB} | \ll 1,
\label{Eq:MainAssumptions}
\end{equation}
where $\hat{g} = d\vartheta^2 + \sin^2\vartheta d\varphi^2$ is the
standard metric on $S^2$, $r$ is the areal radius, and $\tilde{g}$ is
a pseudo-Riemannian metric on $\tilde{M}$ representing the radial part
of the background Schwarzschild metric. In standard Schwarzschild
coordinates $(t,r)$, we have $\tilde{g} = -N(r) dt^2 + N(r)^{-1}
dr^2$, where $N = 1 - 2M/r$. However, we do not assume that $(x^a)$
are Schwarzschild coordinates: with respect to the coordinates $(x^a)$
on $\tilde{M}$, the components of $\tilde{g}$ are given by the
coordinate transformation of the $(t,r)$-components of $\tilde{g}$
which minimizes $|g_{ab}^{(full)} - \tilde{g}_{ab}|$. If the boundary
is spherical, a natural representation of the spacetime manifold in
the neighborhood of the outer boundary as $\tilde{M} \times S^2$
exists, and our hope is that one can enforce clever boundary
conditions on the gauge degrees of freedom guaranteeing that
conditions (\ref{Eq:MainAssumptions}) are satisfied. In this case, a
method for computing $r$ and $\tilde{g}$ from the full metric is the
following \cite{ePeDaNcPeSmT07}: if $\sigma_{ab} = g^{(full)}_{ab}$
and $\gamma_{AB} = g^{(full)}_{AB}$ denote the induced metrics on
$\tilde{\cal M}$ and $S^2$, respectively, then $r$ is defined by $4\pi
r^2 = \int_{S^2} \sqrt{|\gamma|}\sin(\vartheta) d\vartheta d\varphi$
and $\tilde{g} := \int_{S^2} \sigma \sin(\vartheta) d\vartheta
d\varphi$. If the boundary is not spherical, $\tilde{g}$ and $r$ need
to be computed by other methods: a possible approach is to use a
minimization technique in order to optimize the inequalities in
(\ref{Eq:MainAssumptions}).

Having identified $\tilde{M}$, $\tilde{g}$ and $r$, one introduces the
coordinates $t$ and $r_*$ on $\tilde{\cal M}$ which correspond to the
Schwarzschild time and tortoise coordinates, respectively, but which
are defined in a completely geometric way (see Section
\ref{Sect:MetricPert}) so that different observers on $\tilde{M}$
agree on their definition. Consequently, the boundary condition ${\cal
B}_L$ does not depend on the foliation of $\tilde{M}$. It reads
\begin{equation}
{\cal B}_L:~~~~~\left. \partial_t
\left[ r^2(\partial_t + \partial_{r_*})\right]^{L-1}
\left( r^5\Psi_0 \right) \right|_{\partial\Sigma} = 0,
\label{Eq:ImprovedBCI}
\end{equation}
where the Weyl scalar $\Psi_0$ is defined in terms of the
Newman-Penrose null tetrad given in Sect. \ref{Sect:MetricPert}.
Notice that ${\cal B}_1$ is the freezing $\Psi_0$ boundary condition.

After formulating ${\cal B}_L$, we concentrate on improving ${\cal
B}_2$ by reducing spurious reflections off the outer boundary due to
curvature corrections and backscatter. We accomplish this by computing
in- and outgoing quadrupolar solutions for even and odd-parity
perturbations of the Schwarzschild metric $\tilde{g}+r^2\hat{g}$, to
first order in $2M/R$. Using these approximate solutions, we compute
reflection coefficients which quantify the amount of spurious
reflections. When calculating the reflection coefficients, we assume
for simplicity that the outer boundary is an approximate metric sphere
with constant area $4\pi R^2$. This allows us to compare the amount of
reflections resulting from our improved boundary conditions with those
resulting from ${\cal B}_1$, for a spherical outer boundary. Our
boundary condition ${\cal B}_2$ already gives a $1.2(M/R)(kR)^{-1}$
times reduction in the amount of spurious reflections compared to
${\cal B}_1$ for $k R > 1.05$.\footnote{In our previous article
\cite{lBoS06}, we found only an $(M/R)$ times reduction in the amount
of spurious reflections compared to ${\cal B}_1$ for $k R >
1.05$. This discrepancy is due to the fact that the calculations in
\cite{lBoS06} were performed with a slightly different version
$\tilde{\cal B}_2$ of ${\cal B}_2$.}  We find we can improve ${\cal
B}_2$, however, by applying the operator in Eq. (\ref{Eq:ImprovedBCI})
to $r^5 N^{-1}\Psi_0$ rather than to $r^5\Psi_0$ (recall $N = 1-2M/r$)
giving
\begin{equation}
{\cal C}_2:~~~~~\left. \partial_t(\partial_t + \partial_{r_*})
\left( r^5 N^{-1} \Psi_0 \right) \right|_{\partial\Sigma} = 0.
\label{Eq:ImprovedBCII}
\end{equation}
For a spherical outer boundary, ${\cal C}_2$ reduces spurious
reflections due to curvature corrections and backscatter by a factor
of $(15M/2R)(kR)^{-2}$ compared to ${\cal B}_1$, for $k R > 1$. We
have not calculated the reflection coefficients for boundary shapes
other than spherical; however, we stress that our boundary conditions
${\cal B}_L$ and ${\cal C}_2$ can be applied to general outer boundary
shapes. Furthermore, we show that in general, $C_2$ is perfectly
absorbing to first order in $2M/R$ if backscatter is neglected but the
curvature correction terms are taken into account.

Finally, we improve ${\cal C}_2$ so that it is perfectly absorbing up
to order $2M/R$ {\em including backscatter} for quadrupolar
gravitational radiation. This boundary condition reads
\begin{equation}
{\cal D}_2:~~~~~\left[ r(\partial_t + \partial_{r_*})
\left( r^5 N^{-1} \Psi_0 \right)
 + \frac{30M}{r}\int\limits_0^{t/2r}
    \frac{r^5\Psi_0(t - 2r y,r,\vartheta,\varphi) dy}{(1+y)^7} 
 \right]_{\partial\Sigma} = G,
\label{Eq:ImprovedBCIII}
\end{equation}
where $G$ is some appropriate boundary data which can be computed from
the portion of the initial data which is exterior to the outer
boundary (see Eq. (\ref{Eq:B2BackScattering2}) below). In particular,
$G=0$ if the initial data vanish in the region exterior the the outer
boundary.

Our article is organized as follows. In Sect. \ref{Sect:MetricPert},
first we consider an arbitrary spherically symmetric background
manifold of the form $\tilde{\cal M}\times S^2$ and show how, in
vacuum, the coordinates $t$ and $r_*$ can be defined in a geometric
way. This definition is important as it makes it possible to apply our
boundary conditions to quite general foliations of the spacetime and
to spherical as well as non-spherical outer boundary shapes. Next,
using the gauge-invariant perturbation formalism presented in
\cite{oSmT01}, we derive the linearized Weyl scalars $\delta\Psi_0$
and $\delta\Psi_4$ when there are both even and odd-parity
perturbations on a Schwarzschild background in arbitrary spherically
symmetric coordinates. These scalars are invariant with respect to
infinitesimal coordinate transformations and tetrad rotations, and we
use $\delta\Psi_0$ to construct our new boundary conditions. In
Sect. \ref{Sect:InAndOut}, we compute the in- and outgoing quadrupolar
solutions to the Regge-Wheeler and Zerilli equations up to order
$2M/R$. Here, we find the interesting result that for the perturbed
Weyl scalar $\delta\Psi_0$, the ingoing solutions are identical for
both the even and odd-parity sectors, whereas the outgoing solutions
for these sectors differ by a sign in a first order correction
term. In Sect. \ref{Sect:AbsorbingBC}, we discuss the boundary
conditions ${\cal B}_2$ and ${\cal C}_2$ and calculate the
corresponding reflection coefficients up to first order corrections in
$2M/R$. For ${\cal B}_2$, the reflection coefficient decays as
$(2M/R)(kR)^{-5}$ and for ${\cal C}_2$, it decays as $(2M/R)(kR)^{-6}$
(for $kR \gg 1$). Finally, we construct the boundary condition ${\cal
D}_2$, which is perfectly absorbing including backscatter and
curvature correction terms, to first order in $2M/R$. Conclusions are
drawn in Sect. \ref{Sect:Conclusions} where we also present the
boundary condition ${\cal C}_L$ \cite{jB07a}, a generalization of
${\cal C}_2$ derived from the results in \cite{jBwP73}. In the
appendix, we analyze the stability of the boundary condition ${\cal
D}_2$ for a related model problem.

\section{Linear perturbations of a spherically symmetric vacuum spacetime}
\label{Sect:MetricPert}

In this section, we first consider an arbitrary spherically symmetric
background spacetime which can be represented as the product of a
two-manifold $\tilde{\cal M}$ and a two-sphere $S^2$ and show how the
coordinates $t$ and $r_*$ on $\tilde{M}$ can be defined geometrically,
provided that the background spacetime satisfies Einstein's vacuum
equations. We then use the perturbation formalism in \cite{oSmT01},
which is a generalization of the Regge-Wheeler \cite{tRjW57} and
Zerilli \cite{fZ70} equations to arbitrary spherically symmetric
coordinates, to derive formulas for calculating the linearized Weyl
scalars $\delta\Psi_0$, $\delta\Psi_4$, and $\im(\delta\Psi_2)$. As
constructed in this section, these Weyl scalars are gauge-invariants
in the sense that they are invariant with respect to infinitesimal
coordinate transformations and tetrad rotations.

The generalized Regge-Wheeler and Zerilli equations found in
\cite{oSmT01} have the form of wave equations with potentials for the
gauge-invariant scalars $\Phi^{(-)}_{\ell m}$ and $\Phi^{(+)}_{\ell
m}$ which describe, respectively, odd and even-parity linear
fluctuations of the background geometry with angular momentum numbers
$\ell m$. In \cite{oSmT01}, it was also shown that the complete set of
metric perturbations in the Regge-Wheeler gauge can be reconstructed
from the scalars $\Phi^{(-)}_{\ell m}$ and $\Phi^{(+)}_{\ell m}$ {\em
without solving additional differential equations}. In particular, the
linearized Weyl scalars $\delta\Psi_0$, $\delta\Psi_4$ and
$\im(\delta\Psi_2)$, which are gauge-invariant quantities, are
entirely determined by these scalars.

\subsection{Background equations}
\label{SubSect:MetricPertBackground}

We consider a spherically symmetric background of the $2+2$ form
$\tilde{M} \times S^2$ with metric
\begin{equation}
g = \tilde{g}_{ab}\, dx^a dx^b + r^2\,\hat{g}_{AB}\, dx^A dx^B ,
\label{Eq:SphSymMetric}
\end{equation}
where $\hat{g} = d\vartheta^2 + \sin^2\vartheta d\varphi^2$ is the
standard metric on $S^2$ and $\tilde{g}$ and $r$ denote the metric
tensor and a positive function, respectively, defined on the
two-dimensional pseudo-Riemannian orbit space $\tilde{M}$. In what
follows, lower-case Latin indices refer to coordinates on $\tilde{M}$,
while capital Latin indices refer to the coordinates $\vartheta$ and
$\varphi$ on $S^2$. The $ab$ components of the Einstein tensor
belonging to the metric (\ref{Eq:SphSymMetric}) are
\begin{equation}
G_{ab} = - \frac{2}{r}\tilde{\nabla}_a \tilde{\nabla}_b r 
       + \frac{1}{r^2}\tilde{g}_{ab}\left( 2r\tilde{\Delta} r + N - 1 \right),
\label{Eq:Gab}
\end{equation}
where $\tilde{\nabla}$ denotes the covariant derivative compatible
with $\tilde{g}$, $\tilde{\Delta} \equiv
\tilde{\nabla}^a\tilde{\nabla}_a$ and $N = \tilde{g}(dr,dr) =
\tilde{g}^{ab}\tilde{\nabla}_a r\cdot \tilde{\nabla}_b r$. A
coordinate-invariant definition of the ADM mass is given by
\begin{equation}
M = \frac{r}{2}(1 - N).
\label{Eq:DefMass}
\end{equation}
In vacuum, Eq. (\ref{Eq:Gab}) yields
\begin{equation}
\tilde{\nabla}_a\tilde{\nabla}_b r = \frac{M}{r^2}\tilde{g}_{ab}\; .
\label{Eq:BackgroundMainEq}
\end{equation}
Eqs. (\ref{Eq:DefMass}) and (\ref{Eq:BackgroundMainEq}) imply that $M$
is constant.

For the following, we assume that $N > 0$. Since in this article we
are interested in the asymptotic regime of the Schwarzschild
spacetime, where $N$ is close to one, this assumption poses no
problems. The two vector fields
\begin{equation}
t^a = -\tilde{\varepsilon}^{ab}\tilde{\nabla}_b r, \qquad
r^a = \tilde{\nabla}^a r
\end{equation}
on $\tilde{M}$, where $\tilde{\varepsilon}_{ab}$ denotes the natural
volume element on $\tilde{M}$, are orthogonal to each other and are
linearly independent since $r_a r^a = -t_a t^a = N > 0$. As a
consequence of Eq. (\ref{Eq:BackgroundMainEq}), $\tilde{\nabla}_a t_b
= M\tilde{\varepsilon}_{ab}/r^2$; hence, $t^a$ is a Killing vector
field. Furthermore, it follows from (\ref{Eq:BackgroundMainEq}) that
$t^a$ and $r^a$ commute. Therefore, it is possible to introduce
coordinates $(t,r_*)$ on $\tilde{M}$ such that $t^a\partial_a =
\partial_t$ and $r^a\partial_a = \partial_{r_*}$. The coordinates $t$
and $r_*$ correspond to the standard Schwarzschild time and tortoise
coordinates, respectively. For later use, we also introduce the two
null vector fields
\begin{eqnarray}
l^a\partial_a &=& \frac{1}{\sqrt{2N}} \left( t^a + r^a \right)\partial_a
 = \frac{1}{\sqrt{2N}} (\partial_t + \partial_{r_*}),
\label{Eq:SphSymTetradl}\\
k^a\partial_a &=& \frac{1}{\sqrt{2N}} \left( t^a - r^a \right)\partial_a
 = \frac{1}{\sqrt{2N}} (\partial_t - \partial_{r_*}),
\label{Eq:SphSymTetradk}
\end{eqnarray}
satisfying $\tilde{g}_{ab} = -l_a k_b - k_a l_b$ and
$\tilde{\varepsilon}_{ab} = l_a k_b - k_a l_b$. They can be completed
to form a Newman-Penrose null tetrad $\{ l^a, k^a, m^B, \bar{m}^B \}$,
where $m^B$ is a complex null vector field and $\bar{m}^B$ its complex
conjugate such that $r^2\hat{g}_{AB} = m_A \bar{m}_B + \bar{m}_A m_B$
and the volume element $\hat{\varepsilon}$ belonging to $\hat{g}$ is
given by $r^2\hat{\varepsilon}_{AB} = i(m_A\bar{m}_B - \bar{m}_A
m_B)$. The Weyl scalars are defined as
\begin{eqnarray}
\Psi_0 &=& C_{\alpha\beta\mu\nu} l^\alpha m^\beta l^\mu m^\nu,
\nonumber\\
\Psi_1 &=& C_{\alpha\beta\mu\nu} l^\alpha k^\beta l^\mu m^\nu,
\nonumber\\
\Psi_2 &=& C_{\alpha\beta\mu\nu} l^\alpha m^\beta \bar{m}^\mu k^\nu,
\nonumber\\
\Psi_3 &=& C_{\alpha\beta\mu\nu} l^\alpha k^\beta k^\mu\bar{m}^\nu,
\nonumber\\
\Psi_4 &=& C_{\alpha\beta\mu\nu} k^\alpha \bar{m}^\beta k^\mu \bar{m}^\nu,
\nonumber
\end{eqnarray}
where $C_{\alpha\beta\mu\nu}$ denotes the Weyl tensor. For a
spherically symmetric spacetime of the form (\ref{Eq:SphSymMetric})
with the tetrad choice
(\ref{Eq:SphSymTetradl},\ref{Eq:SphSymTetradk}), all the Weyl scalars
vanish except for
\begin{displaymath}
\Psi_2 = -\frac{M}{r^3}\; .
\end{displaymath}

\subsection{Perturbation equations}

Next, let us consider a generic linear perturbation of the background
metric:
\begin{displaymath}
\delta g_{ab} = L_{ab}\; ,\qquad
\delta g_{aB} = Q_{aB}\; ,\qquad
\delta g_{AB} = r^2 K_{AB}\; ,
\end{displaymath}
where the quantities $L_{ab}$, $Q_{aB}$ and $K_{AB}$ depend on the
coordinates $(x^a,x^B)$. With respect to an infinitesimal coordinate
transformation $\delta x^a \mapsto \delta x^a + \xi^a$, $\delta x^B
\mapsto \delta x^B + \eta^B$, these quantities transform according to
\begin{eqnarray}
L_{ab} &\mapsto& L_{ab} + 2\tilde{\nabla}_{(a} \xi_{b)}\; ,
\label{Eq:CoordTransf1}\\
Q_{aB} &\mapsto& Q_{aB} + \hat{\nabla}_B\xi_a 
 + r^2\tilde{\nabla}_a \hat{\eta}_B\, ,
\label{Eq:CoordTransf2}\\
K_{AB} &\mapsto& K_{AB} + \frac{2}{r}\,\hat{g}_{AB}\, r^a \xi_a
 + 2\hat{\nabla}_{(A} \hat{\eta}_{B)}\, ,
\label{Eq:CoordTransf3}
\end{eqnarray}
where $\hat{\nabla}$ denotes the covariant derivative compatible with
$\hat{g}$ and where $\hat{\eta}_A \equiv \hat{g}_{AB}\eta^B$. A
lengthy but straightforward calculation yields the following
expressions for the perturbed Weyl scalars $\delta\Psi_0$ and
$\delta\Psi_4$:
\begin{equation}
\delta\Psi_0 = V_{abCD} l^a l^b m^C m^D, \qquad
\delta\Psi_4 = V_{abCD} k^a k^b \bar{m}^C\bar{m}^D,
\label{Eq:DeltaPsi0Psi4}
\end{equation}
where $V_{abCD}$ is given by
\begin{equation}
V_{abCD} = \frac{1}{2}\left( 
 -\hat{\nabla}_C\hat{\nabla}_D L_{ab}
 + \hat{\nabla}_C\tilde{\nabla}_{(a} Q_{b)D}
 + \hat{\nabla}_D\tilde{\nabla}_{(a} Q_{b)C}
 - \tilde{\nabla}_{(a} r^2 \tilde{\nabla}_{b)} K_{CD} \right)^{tf}
\label{Eq:VabCD}
\end{equation}
and where the superscript $tf$ refers to the trace-free part with
respect to the metric $\hat{g}_{CD}$. It can be verified readily that
$V_{abCD}$ is invariant with respect to the infinitesimal coordinate
transformations given in Eqs. 
(\ref{Eq:CoordTransf1},\ref{Eq:CoordTransf2},\ref{Eq:CoordTransf3}).
Furthermore, since $\delta\Psi_0$ and $\delta\Psi_4$ do not depend on
the perturbed tetrad components, they are also invariant with respect
to infinitesimal tetrad rotations. In fact, this invariance of
$\delta\Psi_0$ and $\delta\Psi_4$ holds for any type D background
metric \cite{Teukolsky73}.

Since the background metric is spherically symmetric, it is convenient
to expand the fields in spherical harmonics. In the resulting
perturbation equations, pieces belonging to different angular momentum
numbers $\ell$ and $m$ decouple. Thus, it is sufficient to consider
one fixed value of $\ell$ and $m$ at a time. The decomposition of the
fields $L_{ab}$, $Q_{aB}$ and $K_{AB}$ into tensor spherical harmonics
reads
\begin{eqnarray}
L_{ab} &=& H_{ab} Y,
\nonumber\\
Q_{aB} &=& Q_a\hat{\nabla}_B Y + h_a\hat{S}_B\,,
\label{Eq:SphDecompQ}\\
K_{AB} &=& K\hat{g}_{AB} Y 
 + G\left[ \hat{\nabla}_A\hat{\nabla}_B\right]^{tf} Y
 + 2k\hat{\nabla}_{(A} \hat{S}_{B)}\, ,
\nonumber
\end{eqnarray}
where $Y = Y^{\ell m}(\vartheta,\varphi)$ denotes the standard
spherical harmonics and $\hat{S}_A
\equiv\hat{\varepsilon}_A{}^B\hat{\nabla}_B Y$. With respect to this,
we obtain
\begin{equation}
V_{abCD} = -\frac{1}{2} H_{ab}^{(inv)} 
            \left[ \hat{\nabla}_C\hat{\nabla}_D\right]^{tf} Y
  + \tilde{\nabla}_{(a} h_{b)}^{(inv)} \hat{\nabla}_{(C}\hat{S}_{D)}\, ,
\end{equation}
where $H_{ab}^{(inv)}$ and $h_b^{(inv)}$ are the gauge-invariant
amplitudes defined in \cite{oSmT01} which, in the Regge-Wheeler gauge,
reduce to the quantities $H_{ab}$ and $h_b$, respectively. These
gauge-invariants were first given in \cite{uGuS79}. For vacuum
perturbations, these gauge-invariant quantities are entirely
determined by two scalar fields $\Phi^{(+)}$ and $\Phi^{(-)}$
according to
\begin{eqnarray}
H_{ab}^{(inv)} &=& -\frac{2}{\lambda + a_0}
 \left( \tilde{\nabla}_a\tilde{\nabla}_b 
         - \frac{1}{2}\tilde{g}_{ab}\tilde{\Delta} \right)
 \left[ r(\lambda + a_0)\Phi^{(+)} \right],
\label{Eq:Habinv}\\
h_b^{(inv)} &=& \tilde{\varepsilon}_{ab}\tilde{\nabla}^a 
  \left[ r\Phi^{(-)} \right],
\label{Eq:hbinv}
\end{eqnarray}
where $\lambda \equiv (\ell-1)(\ell+2)$ and $a_0 \equiv 6M/r$. The two
scalar fields $\Phi^{(+)}$ and $\Phi^{(-)}$ satisfy the Zerilli
\cite{fZ70} and Regge-Wheeler \cite{tRjW57} equations, respectively:
\begin{equation}
\left[ -\tilde{\nabla}^a\tilde{\nabla}_a + V_{\pm}(r) \right] 
 \Phi^{(\pm)} = 0,
\label{Eq:RWZ}
\end{equation}
where the potentials are given by
\begin{eqnarray}
V_{+}(r) &=& \frac{\lambda^2 r^2\left[ \ell(\ell+1)r + 6M \right] 
 + 36M^2(\lambda r + 2M)}{(\lambda r + 6M)^2 r^3}\, ,
\nonumber\\
V_{-}(r) &=& \frac{\ell(\ell+1)}{r^2} - \frac{6M}{r^3}\; .
\nonumber
\end{eqnarray}
Taking everything together, we obtain the following expressions for
the perturbed Weyl scalars $\delta\Psi_0$ and $\delta\Psi_4$:
\begin{eqnarray}
\delta\Psi_0 &=& \frac{1}{r^2}\left\{
 \frac{1}{\lambda + a_0} l^a l^b\tilde{\nabla}_a\tilde{\nabla}_b
 \left[ r(\lambda + a_0)\Phi^{(+)} \right] 
 + i l^a l^b\tilde{\nabla}_a\tilde{\nabla}_b \left[ r\Phi^{(-)} \right] 
 \right\} \hat{m}^C \hat{m}^D\hat{\nabla}_C\hat{\nabla}_D Y,
\label{Eq:Psi0Pert}\\
\delta\Psi_4 &=& \frac{1}{r^2}\left\{
 \frac{1}{\lambda + a_0} k^a k^b\tilde{\nabla}_a\tilde{\nabla}_b
 \left[ r(\lambda + a_0)\Phi^{(+)} \right] 
 + i k^a k^b\tilde{\nabla}_a\tilde{\nabla}_b \left[ r\Phi^{(-)} \right] 
 \right\} \bar{\hat{m}}^C \bar{\hat{m}}^D\hat{\nabla}_C\hat{\nabla}_D Y,
\label{Eq:Psi4Pert}
\end{eqnarray}
where we have defined $\hat{m}^A = r\, m^A$, and where the scalar
fields $\Phi^{(\pm)}$ satisfy Eq. (\ref{Eq:RWZ}).

Finally, we compute the imaginary part of the perturbed Weyl scalar
$\delta\Psi_2$. One first finds
\begin{equation}
\im(\delta\Psi_2) = -\im\left( W_{abCD} l^a k^b m^C\bar{m}^D \right),
\end{equation}
where the quantity
\begin{equation}
 W_{abCD} = \frac{r^2}{2}\left[ 
   \hat{\nabla}_C\tilde{\nabla}_{[a}\left( \frac{Q_{b]D}}{r^2} \right)
 - \hat{\nabla}_D\tilde{\nabla}_{[a}\left( \frac{Q_{b]C}}{r^2} \right)
\right]
\end{equation}
is invariant with respect to the infinitesimal coordinate
transformations given in Eqs.
(\ref{Eq:CoordTransf1},\ref{Eq:CoordTransf2},\ref{Eq:CoordTransf3}).
Next, application of the decomposition (\ref{Eq:SphDecompQ}) yields
\begin{displaymath}
W_{abCD} = r^2\tilde{\nabla}_{[a}\left( \frac{h^{(inv)}_{b]}}{r^2} \right)
 \hat{\nabla}_{[C}\hat{S}_{D]}\, .
\end{displaymath}
Using Eq. (\ref{Eq:hbinv}) and $2r^2
m^C\bar{m}^D\hat{\nabla}_{[C}\hat{S}_{D]} = i
\hat{g}^{CD}\hat{\nabla}_C\hat{\nabla}_D Y$, we then obtain
\begin{displaymath}
\im(\delta\Psi_2) = -\frac{\ell(\ell+1)}{4r}
 \left[ \tilde{\nabla}^a\tilde{\nabla}_a\Phi^{(-)}
 - \frac{2}{r^2}\left( 1 - \frac{3M}{r} \right)\Phi^{(-)} \right] Y.
\end{displaymath}
Finally, we substitute the Regge-Wheeler equation (\ref{Eq:RWZ}) into
the above expression to get
\begin{equation}
\im(\delta\Psi_2) = -\frac{(\ell-1)\ell(\ell+1)(\ell+2)}{4r^3}\Phi^{(-)} Y.
\label{Eq:Psi2Pert}
\end{equation}
Therefore, the gauge-invariant scalar $\Phi^{(-)}$ obeying the
Regge-Wheeler equation can be interpreted as the radial part of the
imaginary part of the perturbed Weyl scalar $\delta\Psi_2$
\cite{rP72b,lBoS06}. Comparison of Eq. (\ref{Eq:Psi2Pert}) with the
quantity $\phi = r^2 h_0$ defined in \cite{lBoS06} gives the relation
$2\phi = (\ell-1)\ell(\ell+1)(\ell+2)\Phi^{(-)}$. It would be nice to
know what the geometric interpretation of the gauge-invariant scalar
$\Phi^{(+)}$ obeying the Zerilli equation is.

\section{In- and outgoing quadrupolar wave solutions}
\label{Sect:InAndOut}

In this section, we compute in- and outgoing solutions to the Zerilli
and Regge-Wheeler equations. Since we are interested only in the
asymptotic regime $r \simeq R$, with $R$ large compared to $M$, we
compute these solutions to first order in $2M/R$. A method for
performing this task is described in \cite{lBoS06}. We start by
computing the outgoing solutions. For this purpose, we introduce the
outgoing Eddington-Finkelstein coordinates $(\tau,\rho)$, which are
related to the geometrically defined coordinate $t$ and the areal
radius $r$ of Sect. \ref{SubSect:MetricPertBackground} through
\begin{displaymath}
\tau = t + r - r_*\, , \qquad
\rho = r,
\end{displaymath}
where the tortoise coordinate $r_*$ is defined by $r_*(r) = r - 4M +
2M\log ( r/2M - 1)$. In these coordinates, the two-metric reads
\begin{displaymath}
\tilde{g} = -d\tau^2 + d\rho^2 + \frac{2M}{\rho} (d\tau - d\rho)^2.
\end{displaymath}
With respect to this metric, the Zerilli and Regge-Wheeler equations
(\ref{Eq:RWZ}) can be written as
\begin{equation}
\left[ \partial_\tau^2 - \partial_\rho^2 + \frac{\ell(\ell+1)}{\rho^2} \right]
\Phi^{(\pm)} = -\frac{2M}{\rho}\left[ (\partial_\tau + \partial_\rho)^2
 - \frac{1}{\rho}(\partial_\tau + \partial_\rho) - \frac{3a_{\pm}}{\rho^2}
 + O\left( \frac{2M}{R} \right) \right] \Phi^{(\pm)},
\end{equation}
where $a_+ = (\lambda + 4)/\lambda$ and $a_- = 1$. In order to compute
the solution to first order in $2M/R$, we make the ansatz (omitting the
superscript $(\pm)$ on $\Phi$)
\begin{displaymath}
\Phi(\tau,\rho) = \Phi_0(\tau,\rho) + \frac{2M}{R} g_1(\tau,\rho),
\end{displaymath}
where $\Phi_0$ is a solution of the Regge-Wheeler-Zerilli equations
(\ref{Eq:RWZ}) for $2M/R = 0$, and where the function $g_1$ satisfies
the equation
\begin{equation}
\left[ \partial_\tau^2 - \partial_\rho^2 + \frac{\ell(\ell+1)}{\rho^2} \right]
g_1(\tau,\rho) = -\frac{R}{\rho}\left[ (\partial_\tau + \partial_\rho)^2
 - \frac{1}{\rho}(\partial_\tau + \partial_\rho) - \frac{3a_{\pm}}{\rho^2}
 \right] \Phi_0\, .
\label{Eq:g1}
\end{equation}
In the following, we compute $\Phi$ for quadrupolar radiation
($\ell=2$). In this case, the outgoing flat background solution
$\Phi_0$ is given by
\begin{displaymath}
\Phi_0(\tau,\rho) = U^{(2)}(\rho-\tau) - \frac{3}{\rho} U^{(1)}(\rho-\tau)
 + \frac{3}{\rho^2} U(\rho-\tau),
\end{displaymath}
where $U$ is a smooth function on the real axis with $k$'th derivative
$U^{(k)}$. In the following, we also assume that $U$ is bounded and
that it vanishes for negative enough values of its argument. The
corresponding solution for $g_1$ is given by
\begin{equation}
g_1(\tau,\rho) = \frac{3(1-a_\pm)R}{2\rho^3} U(\rho-\tau)
 + \frac{3a_\pm R}{4\rho^2} U^{(1)}(\rho-\tau)
 + \frac{R}{4}\int\limits_{\rho-\tau}^\infty K_2(\tau,\rho,x) U(x) dx,
\label{Eq:FirstOrderCorrection}
\end{equation}
where the integral kernel $K_2$ reads
\begin{displaymath}
K_2(\tau,\rho,x) = \frac{3}{2\rho^4}\left[ w^{-4} + 2w^{-3} + 2w^{-2} 
 \right]_{w = \frac{\tau+\rho+x}{2\rho}}\; ,\qquad x \geq \rho - \tau,
\end{displaymath}
and satisfies
\begin{displaymath}
\left[ \partial_{\tau}^2 - \partial_{\rho}^2 
  + \frac{6}{\rho^2} \right] K_2(\tau,\rho,x) = 0, \qquad
K_2(\tau,\rho,\rho-\tau) = \frac{15}{2\rho^4}\; .
\end{displaymath}
The solution $g_1$ is not unique; one can add an arbitrary homogeneous
solution of Eq. (\ref{Eq:g1}) to it. However, $g_1$ is uniquely
characterized by the following conditions:
\begin{eqnarray}
\lim\limits_{\rho\to\infty} g_1(const.+\rho,\rho) = 0 &&
\hbox{($g_1$ vanishes at future null infinity)},
\nonumber\\
\lim\limits_{\rho\to\infty} g_1(const.-\rho,\rho) = 0 &&
\hbox{($g_1$ vanishes at past null infinity)},
\nonumber\\
\lim\limits_{\tau\to\infty} g_1(\tau,const.)~~~~~ = 0 &&
\hbox{($g_1$ vanishes at future time-like infinity)}.
\nonumber
\end{eqnarray}
Notice that the third requirement is necessary in order to exclude
solutions of the form $g_1(\tau,\rho) = (c_1 \tau + c_0)\rho^{-2}$ (where
$c_0$ and $c_1$ are some constants) of the zero mass
Regge-Wheeler-Zerilli equations (\ref{Eq:RWZ}), which vanish at both
future and past null infinity.

Summarizing, we have the following outgoing approximate solutions of
the Regge-Wheeler-Zerilli equations for $\ell=2$:
\begin{eqnarray}
\Phi^{(\pm)}_{\nearrow}(t,r) &=& 
 U_\pm^{(2)}(r_*-t) - \frac{3}{r} U_\pm^{(1)}(r_*-t) 
 + \frac{3}{r^2} U_\pm(r_*-t)
\nonumber\\
 &+& \frac{2M}{R}\left[  \frac{3a_\pm R}{4 r^2} U_\pm^{(1)}(r_* - t)
  + \frac{3(1-a_\pm)R}{2r^3} U_\pm(r_* - t)
  + \frac{R}{4}\int\limits_{r_* - t}^\infty K_2(r-r_*+t,r,x) U_\pm(x) dx 
 \right] + O\left( \frac{2M}{R} \right)^2,
\nonumber
\end{eqnarray}
where $a_+ = 2$, $a_- = 1$. Since the Regge-Wheeler-Zerilli equations
(\ref{Eq:RWZ}) are time-symmetric, corresponding ingoing solutions are
obtained from this by merely flipping the sign of $t$:
\begin{eqnarray}
\Phi^{(\pm)}_{\nwarrow}(t,r) &=&
 V_\pm^{(2)}(r_*+t) - \frac{3}{r} V_\pm^{(1)}(r_*+t) 
  + \frac{3}{r^2} V_\pm(r_*+t)
\nonumber\\
 &+& \frac{2M}{R}\left[  \frac{3a_\pm R}{4 r^2} V_\pm^{(1)}(r_* + t)
  + \frac{3(1-a_\pm)R}{2r^3} V_\pm(r_* + t)
  + \frac{R}{4}\int\limits_{r_* + t}^\infty K_2(r-r_*-t,r,x) V_\pm(x) dx 
 \right] + O\left( \frac{2M}{R} \right)^2.
\nonumber
\end{eqnarray}
Using equation (\ref{Eq:Psi0Pert}) and the identity
\begin{displaymath}
\frac{1}{r} l^a l^b\tilde{\nabla}_a\tilde{\nabla}_b(r\Phi) 
 = \frac{N}{2r^4} \left[ \frac{r^2}{N}(\partial_t + \partial_{r_*}) \right]^2
   \Phi,
\end{displaymath}
which holds for the Schwarzschild background, we obtain the following
expressions for the perturbed Weyl scalar $\delta\Psi_0$:
\begin{equation}
\delta\Psi_0 = \frac{1}{r}\left( \psi^{(+)}_0 + i\,\psi^{(-)}_0 \right)
 \hat{m}^C \hat{m}^D\hat{\nabla}_C\hat{\nabla}_D Y,
\label{Eq:DeltaPsi0Decomp}
\end{equation}
where
\begin{eqnarray}
\psi^{(\pm)}_{0\nearrow}(t,r) &=& \frac{3}{r^4 N}\left\{ U_\pm(r_* - t)
 + \frac{2M}{r}\left[ -2U_\pm(r_* - t) \mp \frac{r}{4}U_\pm^{(1)}(r_* - t)
 + \frac{1}{2}\int\limits_0^\infty k(1+y) U_\pm(r_* - t + 2 r y) dy \right] 
 \right.
\nonumber\\
 &+& \left. O\left( \frac{2M}{R} \right)^2 \right\},
\label{Eq:Psi0SchwOut}\\
\psi^{(\pm)}_{0\nwarrow}(t,r) &=& \frac{3}{r^4 N}\left\{ V_\pm(r_* + t)
 - 2r V_\pm^{(1)}(r_* + t) + 2r^2 V_\pm^{(2)}(r_* + t) 
 - \frac{4}{3} r^3 V_\pm^{(3)}(r_* + t) 
 + \frac{2}{3} r^4 V_\pm^{(4)}(r_* + t) \right.
\nonumber\\
 &+& \left. \frac{2M}{r}\left[ \frac{1}{2}r^2 V_\pm^{(2)}(r_* + t) 
 - \frac{1}{2}r^3 V_\pm^{(3)}(r_* + t)
 + \frac{1}{2}\int\limits_0^\infty \frac{V_\pm(r_* + t + 2 r y) dy}{(1+y)^2} 
\right] + O\left( \frac{2M}{R} \right)^2 \right\},
\label{Eq:Psi0SchwIn}
\end{eqnarray}
with $k(w) \equiv 5w^{-6} + 4w^{-5} + 3w^{-4} + 2w^{-3} + w^{-2}$. In
the odd-parity case, these expressions agree with the expressions
obtained in \cite{lBoS06} except that $U_-$ and $V_-$ are rescaled by
a factor of $12$. This factor is explained by Eq. (\ref{Eq:Psi2Pert}).
Notice that the expressions in the even $(+)$ and odd $(-)$ case are
exactly identical {\em except for the sign in the third term of
$\psi^{(\pm)}_{0\nearrow}$}. Finally, we notice that
$\delta\Psi_{0\nearrow}$ decays as $r^{-5}$ along the outgoing
future-directed null geodesics while $\delta\Psi_{0\nwarrow}$ decays
as $r^{-1}$ along the outgoing past-directed null geodesics, as
expected from the peeling theorem \cite{rP65}.

The first two terms inside the square brackets in the expressions for
$\psi^{(\pm)}_{0\nearrow}$ and $\psi^{(\pm)}_{0\nwarrow}$ describe the
curvature corrections, while the integral terms describe the
backscatter. Notice that, by construction, the solution given in
Eq. (\ref{Eq:Psi0SchwIn}) does not yield any radiation at future null
infinity and hence the integral in that expression is acausal in the
sense that it involves integrating over the infinite future of
$V_\pm$. However, since we are only needing the zeroth order
contribution in $M/R$ of $\psi^{(\pm)}_{0\nwarrow}$ in this article,
this acausality does not introduce any problems here.

\section{Absorbing boundary conditions}
\label{Sect:AbsorbingBC}

In this section, we focus on the boundary condition ${\cal B}_2$
which, as formulated in this article, is applicable to general outer
boundary shapes and fairly general foliations of the spacetime. We
improve ${\cal B}_2$ to give two new boundary conditions, ${\cal C}_2$
and ${\cal D}_2$, which reduce the amount of spurious reflections due
to curvature and backscatter off of curvature. ${\cal C}_2$ is
obtained by modifying ${\cal B}_2$ so that it is perfectly absorbing
for quadrupolar gravitational radiation up to order $2M/R$ curvature
(but not backscatter) terms.  We find that the reflection coefficient
for ${\cal C}_2$ is smaller than that for ${\cal B}_2$ by a factor of
the order of $kR$, where $k$ is the characteristic wave number of the
gravitational wave. Finally, we construct ${\cal D}_2$, which is
perfectly absorbing up to order $2M/R$ in both curvature and
backscatter for quadrupolar waves. By nature of the backscatter
effect, ${\cal D}_2$ is non-local: as we will see, it involves an
integral over the past boundary.

\subsection{An estimate for the amount of spurious reflections}
\label{Sect:AbsorbingBCa}

\begin{figure}[htb]
\centerline{
\includegraphics[width=10cm]{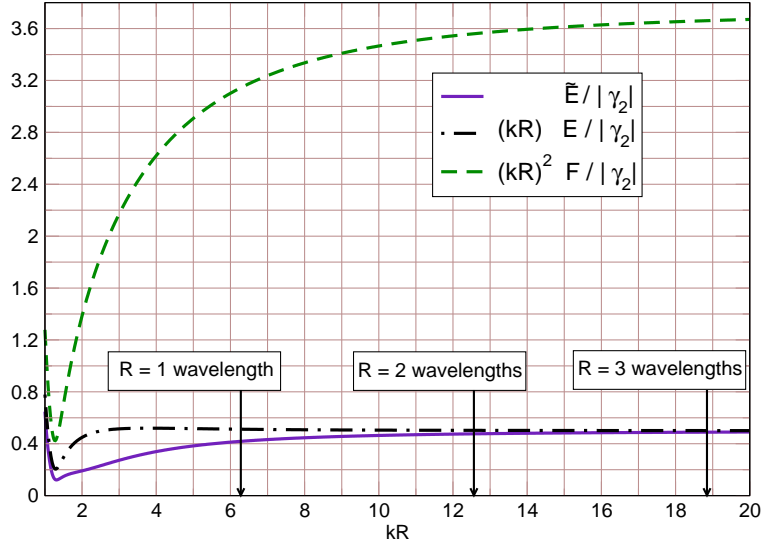}}
\vspace{1cm}
\caption{Comparison of the functions $\tilde{E}(kR)$, $E(kR)$, and
$F(kR)$ (which are the reflection coefficients times $R/(2M)$ for the
boundary conditions $\tilde{\cal B}_2$, ${\cal B}_2$, and ${\cal
C}_2$) divided by the reflection coefficient $|\gamma_2(kR)|$ for the
freezing-$\Psi_0$ boundary condition ${\cal B}_1$, versus $k
R$. Notice that the reflection coefficients are scaled by factors of
$k R$ for the sake of the presentation.}
\label{Fig:ImprovedBCs}
\end{figure}

Here, we estimate the amount of spurious reflections for quadrupolar
waves for the boundary condition ${\cal B}_2$, and compare it to the
corresponding result for the freezing $\Psi_0$ boundary condition
${\cal B}_1$. In order to calculate the reflection coefficients, we
make the following three assumptions:
\begin{enumerate}
\item[(i)] the boundary $\partial\Sigma$ is an approximate metric
two-sphere with constant area $4\pi R^2$,
\item[(ii)] the outer boundary lies in the weak field regime, i.e. $R
\gg 2M$,
\item[(iii)] if $k$ is the characteristic wavenumber of the wave, $M k
\ll 1$.
\end{enumerate}
With assumptions (i) and (ii), we can linearize the field
equations about the exterior Schwarzschild spacetime of mass $M$ near
the outer boundary, where $M$ is the total mass of the system, and
assume that the outer boundary is located at constant areal radius
$r=R$. In- and outgoing quadrupolar wave solutions near the outer
boundary are described by the expressions found in the previous
section. The general solution to the IBVP with the boundary condition
${\cal B}_2$ consists of a superposition of an in- and a outgoing
solution:
\begin{equation}
\psi_0 = \psi_{0\nearrow} + \gamma\, \psi_{0\nwarrow}\; ,
\label{Eq:Superposition}
\end{equation}
with $\psi_{0\nearrow}$ and $\psi_{0\nwarrow}$ given by
Eqs. (\ref{Eq:Psi0SchwOut}) and (\ref{Eq:Psi0SchwIn}), respectively,
and where $\gamma$ is an amplitude reflection coefficient. To
determine $\gamma$, we assume monochromatic radiation,
\begin{equation}
U(r_* - t) = e^{ik(r_* - t)},\qquad
V(r_* + t) = e^{-ik(r_* + t)}
\label{Eq:Monochromatic}
\end{equation}
with $k$ a given wave number satisfying $0 < k \ll 1/M$, and plug the
ansatz (\ref{Eq:Superposition}) into the boundary condition ${\cal
B}_2$. First, we find (the result in the even- and odd-parity case is
exactly identical, so we omit the superscript $(\pm)$)
\begin{eqnarray}
r(\partial_t + \partial_{r_*}) (r^4\psi_{0\nearrow})(t,r)
 &=& \frac{6M}{r} \left[ U(r_* - t) 
  - 15\int\limits_0^\infty \frac{U(r_* - t + 2ry) dy}{(1+y)^7} \right]
  + O\left( \frac{2M}{R} \right)^2
\nonumber\\
r(\partial_t + \partial_{r_*}) (r^4\psi_{0\nwarrow})(t,r)
 &=& 4 r^5 V^{(5)}(r_* + t) + O\left( \frac{2M}{R} \right).
\nonumber
\end{eqnarray}
Using the monochromatic ansatz (\ref{Eq:Monochromatic}) and imposing
${\cal B}_2$, we then find
\begin{displaymath}
\gamma = \gamma\left( kR, \frac{2M}{R} \right)
       = \frac{3M}{2i R} \frac{e^{2ik R_*}}{(kR)^5} 
         \left[ 1 - 15 C_7(kR) - 15i S_7(kR) \right]
       + O\left( \frac{2M}{R} \right)^2
\end{displaymath}
with the integrals
\begin{displaymath}
C_n(z) = \int\limits_0^\infty \frac{\cos(2zy)}{(1+y)^n}\, dy, \qquad
S_n(z) = \int\limits_0^\infty \frac{\sin(2zy)}{(1+y)^n}\, dy, \qquad
n \geq 2,
\end{displaymath}
which have the asymptotic expansions $(2z)^2 C_n(z) = n[1 -
(n+1)(n+2)/(2z)^2 + O(z^{-4})]$, $2z S_n(z) = 1 - n(n+1)/(2z)^2 +
O(z^{-4})$ for $z \gg 1$. Finally, we obtain
\begin{equation}
\Big| \gamma\left( kR,\frac{2M}{R} \right) \Big|
 = \frac{2M}{R} E(k R) + O\left( \frac{2M}{R} \right)^2,
\label{Eq:ReflCoeff}
\end{equation}
where
\begin{displaymath}
E(z) = \frac{3}{4 z^5} \left[ \left( 1 - 15 C_7(z) \right)^2 
 + \left( 15 S_7(z) \right)^2 \right]^{1/2}.
\end{displaymath}
Because of assumption (iii), we may neglect the quadratic corrections
in $2M/R$. (If $k M$ is of the order of unity or larger, powers of $k
R$ which are multiplied by $(2M/R)^2$ might be comparable in size to
or larger than terms of the form $M/R$ times unity. For example, see
the exact outgoing solutions on a Schwarzschild background obtained in
\cite{jBwP73}.) For $k R \gg 1$, the reflection coefficient goes as
$(2M/R)(k R)^{-5}$. Comparing this result with the corresponding
result for ${\cal B}_1$, for which the reflection coefficient
\cite{lBoS06}
\begin{displaymath}
|\gamma_2(kR)| 
 = \left[ 1 - \frac{8}{9}(k R)^6 + \frac{4}{9}(k R)^8 \right]^{-1/2} 
 + O\left( \frac{2M}{R} \right)
\end{displaymath}
decays as $(k R)^{-4}$ for $k R \gg 1$, we see that we gain a
factor of $(2M/R) (k R)^{-1}$ in the decay rate. The function $(kR)
E(kR)/|\gamma_2(kR)|$ is plotted in Figure \ref{Fig:ImprovedBCs}. This
plot, together with the asymptotic expansion $2z E(z)/|\gamma_2(z)| =
1 + 7/(8z^2) + O(z^{-4})$, suggests that for $kR > 1.05$, this function
does not exceed $0.6$. Therefore, if $kR > 1.05$, the boundary
condition ${\cal B}_2$ gives a reflection coefficient which is smaller
by a factor of $1.2M/(kR^2)$ than the one for ${\cal B}_1$.

One can do better: applying the operator $r(\partial_t +
\partial_{r_*})$ to the function $r^5 N^{-1}\Psi_0$ instead of
$r^5\Psi_0$ gives the boundary condition ${\cal C}_2$ given in
Eq. (\ref{Eq:ImprovedBCII}). Since in this case, outgoing quadrupolar
wave solutions satisfy
\begin{equation}
r(\partial_t + \partial_{r_*}) (r^4 N^{-1}\psi_{0\nearrow})(t,r)
 = -\frac{90M}{r}\int\limits_0^\infty \frac{U(r_* - t + 2ry) dy}{(1+y)^7}
 + O\left( \frac{2M}{R} \right)^2,
\label{Eq:DerivPsi0Out}
\end{equation}
the term $6M U(r_* - t)/r$ is eliminated. Physically, this means that
the boundary condition ${\cal C}_2$ takes care of the curvature
correction terms and is perfectly absorbing if the backscatter off the
curvature is neglected. The function $E(z)$ in
Eq. (\ref{Eq:ReflCoeff}) is now replaced by the function
\begin{displaymath}
F(z) = \frac{45}{4z^5} \left[ C_7(z)^2 + S_7(z)^2 \right]^{1/2},
\end{displaymath}
so the new reflection coefficient decays as $(2M/R)(kR)^{-6}$ as
$kR\to\infty$. Therefore, we gain a further power of $(kR)^{-1}$ in
the decay rate! The function $(kR)^2 F(kR)/|\gamma_2(kR)|$ is plotted
in Figure \ref{Fig:ImprovedBCs}. Together with the asymptotic
expansion $4z^2 F(z)/|\gamma_2(z)| = 15[1 - 71/(8z^2) + O(z^{-4})]$,
this plot suggests that for $k R > 1$, $(kR)^2 F(kR)/|\gamma_2(kR)|$
does not exceed $15/4$. Therefore, if $k R > 1$, the improved boundary
condition ${\cal C}_2$ gives a reflection coefficient which is smaller
by a factor of $15M/(2k^2R^3)$ than the one for ${\cal B}_1$.
\vspace{1cm}

For comparison purposes, we plot the function
$\tilde{E}(z)/|\gamma_2(z)|$ in Figure \ref{Fig:ImprovedBCs}, where
the function $\tilde{E}(z)$ was computed in \cite{lBoS06} and
represents the reflection coefficient resulting from the boundary
condition $\tilde{\cal B}_2: \left. \partial_t (\partial_t +
\partial_r)(r^5\Psi_0) \right|_{r=R} = 0$ (the same as ${\cal B}_2$
but with $\partial_{r_*}$ replaced by $\partial_r$). We have checked
that -- as with the functions $E(z)$ and $F(z)$ -- the function
$\tilde{E}(z)$ is exactly the same in the odd- and the even-parity
sector.

\subsection{New boundary conditions which reduce spurious reflections 
due to the backscattering}

Here, we show how to obtain even better boundary conditions: ones
which are {\em perfectly absorbing} up to order $2M/R$ {\em including
the backscatter} for quadrupolar gravitational waves. In other words,
we now construct boundary conditions which have the property of giving
a reflection coefficient that is of the order of $(2M/R)^2$ for
quadrupolar radiation.

First, by comparing Eq. (\ref{Eq:DerivPsi0Out}) with
$\psi_{0\nearrow}(t,r) = 3 r^{-4} \left[ U(r_* - t) + O(2M/R) \right]$,
we notice that the boundary condition
\begin{equation}
\left. r(\partial_t + \partial_{r_*})(r^5 N^{-1}\Psi_0) \right|_{r=R}
 + \frac{30M}{R} \int\limits_0^\infty
   \frac{R^5\Psi_0(t - 2R y,R,\vartheta,\varphi) dy}{(1+y)^7} = 0
\label{Eq:B2BackScattering1}
\end{equation}
is satisfied up to first order in $2M/R$ for the outgoing solution,
the integral operator taking care of the leading order corrections in
the backscatter. However, a problem with the boundary condition in
Eq. (\ref{Eq:B2BackScattering1}) is that it requires knowledge of
$\Psi_0$ on the whole past portion of the boundary surface. So for a
Cauchy formulation which starts an evolution from initial data at the
time slice $t=0$, say, this boundary condition is impractical. In
order to fix this problem, we split the integral in the expression
(\ref{Eq:DerivPsi0Out}) into two parts:
\begin{displaymath}
 \int\limits_0^\infty \frac{U(R_* - t + 2Ry) dy}{(1+y)^7}
 = \int\limits_0^{t/2R} \frac{U(R_* - (t - 2Ry)) dy}{(1+y)^7}
 + \int\limits_{t/2R}^\infty \frac{U(R_* + 2Ry - t) dy}{(1+y)^7}\; .
\end{displaymath}
For a fixed event $(t,R)$ with $t \geq 0$, $R \gg 2M$, the first
integral on the right-hand side can be interpreted as an integral over
the past line $(\tau,R)$, $0 \leq \tau \leq t$, whereas the second
integral can be interpreted as an integral over the line $(0,r)$, $R
\leq r < \infty$. With this in mind, we replace the boundary condition
(\ref{Eq:B2BackScattering1}) with the condition:
\begin{equation}
\left. r(\partial_t + \partial_{r_*})(r^5 N^{-1}\Psi_0) \right|_{r=R} 
 + \frac{30M}{R}
   \int\limits_0^{t/2R}
    \frac{R^5\Psi_0(t - 2R y,R,\vartheta,\varphi) dy}{(1+y)^7} 
 = -\frac{30M}{R}\int\limits_{t/2R}^\infty 
 \left. \frac{s^5\Psi_0(0,s,\vartheta,\varphi)}{(1+y)^7}
 \right|_{s = (R_* + 2Ry - t)_\bullet} dy,
\label{Eq:B2BackScattering2}
\end{equation}
where $(.)_\bullet: (-\infty,\infty) \to (2M,\infty)$ denotes the
inverse of the transformation $(.)_*: (2M,\infty) \to
(-\infty,\infty)$ which maps $r$ to $r_*$. In particular,
$(r_\bullet)_* = r$ and $(r_*)_\bullet = r$. Now the integral on the
left-hand side of Eq. (\ref{Eq:B2BackScattering2}) only involves the
past portion $(\tau,R)$, $0 \leq \tau \leq t$, of the boundary point
$(t,R)$. The past portion $(\tau,R)$, $\tau < 0$, which is not
available in a Cauchy evolution starting at $t=0$, is replaced by an
integral over the exterior $r > R$ of the initial data. If the initial
data are compactly supported in the interval $(0,R)$, this integral is
zero and can be discarded. If the initial data do not vanish for $r >
R$, one uses the exterior part of the initial data in order to define
the integral on the right-hand side of
Eq. (\ref{Eq:B2BackScattering2}), which acts as a source term in the
new boundary condition. The generalization of
Eq. (\ref{Eq:B2BackScattering2}) to non-spherical boundaries yields
the boundary condition ${\cal D}_2$ given in
Eq. (\ref{Eq:ImprovedBCIII}).

Summarizing, we have constructed a new boundary condition which is
perfectly absorbing up to order $2M/R$ in the backscatter for
quadrupolar gravitational radiation. In the appendix, we show for a
related model problem that this boundary condition is stable in the
sense that it admits an energy estimate.

\section{Conclusions}
\label{Sect:Conclusions}

In this article, we have analyzed the problem of specifying absorbing
outer boundary conditions in general relativity. To this end, we have
constructed approximate solutions to the Einstein equations linearized
about a Schwarzschild metric of mass $M$ (where $M$ describes the ADM
mass of the total spacetime), using the generalized
Regge-Wheeler-Zerilli formalism of Ref. \cite{oSmT01}. These solutions
represent in- and outgoing gravitational radiation in the asymptotic
regime $2M/r \ll 1$, where $r$ denotes the areal radius. Since far
enough from the strong field region, any stationary asymptotically
flat spacetime looks like a Schwarzschild spacetime of mass $M$, these
solutions describe, up to order $2M/r$, in- and outgoing gravitational
radiation propagating on the asymptotic region of any such spacetime
background.

There is an interesting difference between the propagation of even and
odd-parity outgoing gravitational linearized radiation on a
Schwarzschild background. This difference can be obtained by computing
the scalars $\psi_0^{(\pm)}$ and $\psi_4^{(\pm)}$ on $\tilde{\cal M}$
in the decomposition
\begin{eqnarray}
l^a l^b V_{abCD} &=& 
   r\psi_0^{(+)} \left[ \hat{\nabla}_C\hat{\nabla}_D \right]^{tf} Y 
 - r\psi_0^{(-)} \hat{\nabla}_{(C} \hat{S}_{D)}\; ,
\nonumber\\
k^a k^b V_{abCD} &=& 
   r\psi_4^{(+)} \left[ \hat{\nabla}_C\hat{\nabla}_D \right]^{tf} Y 
 + r\psi_4^{(-)} \hat{\nabla}_{(C} \hat{S}_{D)}\;
\nonumber
\end{eqnarray}
of the gauge-invariant quantity $V_{abCD}$ defined in
Eq. (\ref{Eq:VabCD}) (recall that $(+)$ refers to even and $(-)$ to
odd parity). Using Eqs. (\ref{Eq:Psi0SchwOut}), (\ref{Eq:Psi4Pert}),
and the identity $2 r^3 k^a k^b\tilde{\nabla}_a\tilde{\nabla}_b(r\Phi)
= N [r^2 N^{-1}(\partial_t - \partial_{r_*})]^2 \Phi$, we obtain the
following expressions for the outgoing solution:
\begin{eqnarray}
\psi^{(\pm)}_{0\nearrow}(t,r) &=& \frac{3}{r^4 N}\left\{ U_\pm(r_* - t)
 + \frac{2M}{r}\left[ -2U_\pm(r_* - t) \mp \frac{r}{4}U_\pm^{(1)}(r_* - t)
 + \frac{1}{2}\int\limits_0^\infty k(1+y) U_\pm(r_* - t + 2 r y) dy \right] 
 \right.
\nonumber\\
 &+& \left. O\left( \frac{2M}{R} \right)^2 \right\},
\label{Eq:Psi0SchwOutBis}\\
\psi^{(\pm)}_{4\nearrow}(t,r) &=& \frac{3}{r^4 N}\left\{ U_\pm(r_* - t)
 - 2r U_\pm^{(1)}(r_* - t) + 2r^2 U_\pm^{(2)}(r_* - t) 
 - \frac{4}{3} r^3 U_\pm^{(3)}(r_* - t) + \frac{2}{3} r^4 U_\pm^{(4)}(r_* - t) 
 \right.
\nonumber\\
 &+& \left. \frac{2M}{r}\left[ \frac{1}{2}r^2 U_\pm^{(2)}(r_* - t) 
 - \frac{1}{2}r^3 U_\pm^{(3)}(r_* - t)
 + \frac{1}{2}\int\limits_0^\infty \frac{U_\pm(r_* - t + 2 r y) dy}{(1+y)^2} 
\right] + O\left( \frac{2M}{R} \right)^2 \right\},
\label{Eq:Psi4SchwOut}
\end{eqnarray}
with $N = 1 - 2M/r$ and $k(w) \equiv 5w^{-6} + 4w^{-5} + 3w^{-4} +
2w^{-3} + w^{-2}$. Taking into account the different normalization of
the null tetrad vectors $l^a$ and $k^a$, and identifying $G(u) = 12[
U_+(r_* - t) + i U_-(r_* - t)]$ with $u = t - r_*$, expression
(\ref{Eq:Psi4SchwOut}) agrees precisely with the expression given in
Eq. (4.26) of Ref. \cite{jBwP73}, up to order $M/R$. However, the
identification in Ref. \cite{jBwP73} of $G(u)$ in Eq. (4.18) with
$G(u)$ in Eq. (4.26) is incorrect. Expression
(\ref{Eq:Psi0SchwOutBis}) agrees with the one given in Eq. (4.18) of
Ref. \cite{jBwP73} with the identification $G(u) = 3[U_+(r_*-t) + i
U_-(r_*-t)] - 3M[ U^{(1)}_+(r_*-t) - i U^{(1)}_-(r_*-t)]/2$. While the
expressions for $\psi^{(\pm)}_{4\nearrow}$ are exactly the same in the
odd and even parity sectors, those for the quantity
$\psi^{(\pm)}_{0\nearrow}$ differ by a sign in the third
term.\footnote{This result contradicts the Teukolsky-Starobinsky
relation given in Eq. (41) of chapter 9 in \cite{Chandrasekhar}, which
does not distinguish between odd- and even-parity perturbations. Based
on the generalized Regge-Wheeler-Zerilli formalism used here, we have
re-derived the Teukolsky-Starobinsky identities for Schwarzschild
perturbations and verified that our result is correct. In particular,
we have calculated the constant ${\cal C}$ in Eq.  (41) of chapter 9
in \cite{Chandrasekhar} and found that it is different in the odd- and
even parity sectors.} For $M = 0$, there is no difference: even- and
odd-parity waves behave in exactly the same way. But as soon as the
curvature of the background is taken into account, there is a symmetry
breaking between the two parity sectors. Provided sufficient accuracy
is available, one should be able to detect this difference in actual
numerical simulations.

Using the ingoing and outgoing solutions, and the assumption that the
outer boundary is an approximate metric sphere of constant area, we
have computed the reflection coefficient which quantifies the amount
of spurious radiation reflected into the computational domain by our
boundary condition ${\cal B}_2$ given in
Eq. (\ref{Eq:ImprovedBCI}). While this boundary condition is perfectly
absorbing for linearized quadrupolar waves, there are small spurious
reflections when the curvature of the background is taken into
account. However, we have shown that the reflection coefficient due to
this effect is {\em extremely small}; it is about a factor $M/(k R^2)$
smaller than the corresponding reflection coefficient for the freezing
$\Psi_0$ boundary condition which was already found to be small in
\cite{lBoS06}. By slightly modifying ${\cal B}_2$, we obtain a new
boundary condition ${\cal C}_2$ which results in even less
reflections. To give a specific example, if the outer boundary is
spherical with areal radius $R > 100M$ and quadrupolar waves with
wavelength $50M$ are considered, the freezing $\Psi_0$ boundary
condition yields a reflection coefficient which is smaller than
$6.5\times 10^{-5}$. If the boundary condition ${\cal C}_2$ is used
instead, the reflection coefficient is smaller than $3.1\times
10^{-8}$. Finally, we have found a boundary condition ${\cal D}_2$
which takes into account the leading order behavior of the backscatter
and is perfectly absorbing up to order $2M/R$ for quadrupolar
gravitational radiation. However, this boundary condition is
non-local: it involves two integral operators. The first operator
involves an integral over the past of the boundary point, the second
an integral over the exterior portion of the initial data. It is shown
in the appendix for a related model problem that such non-local
boundary conditions are stable in that that they admit an energy
estimate.

Although our reflection coefficient calculations assume a spherical
outer boundary, the formulation of our boundary conditions do not: all
that is needed for their construction is that spacetime in the
vicinity of the outer boundary can be represented as the product of a
two-manifold $\tilde{\cal M}$ and a two-sphere $S^2$ such that the
assumptions (\ref{Eq:MainAssumptions}) hold.

We have constructed the boundary conditions ${\cal C}_2$ and ${\cal
D}_2$ to reduce spurious reflections for quadrupolar waves. However,
in many physically interesting scenarios such as the binary black hole
problem, it is possible that when implementing ${\cal D}_2$, spurious
reflections from the $\ell=3$ octupolar contribution to the
gravitational radiation could be as large or larger than the $\ell=2$
quadrupolar backscatter correction. Bardeen \cite{jB07a} has
generalized ${\cal C}_2$ using the results in \cite{jBwP73} to give
${\cal C}_L$, a hierarchy of local boundary conditions which is
perfectly absorbing including curvature corrections (but neglecting
backscatter) to order $2M/R$ for all multipoles of gravitational
radiation up to a given angular momentum number $L$. He obtains
\begin{equation}
{\cal C}_L:~~~~~\left. 
\partial_t~[r^2N^{-1}(\partial_t+\partial_{r_*})]^{L-1}
\left( r^5 N^{-1} \Psi_0 \right) \right|_{\partial\Sigma} = 0.
\label{Eq:C_L}
\end{equation}
To determine whether ${\cal C}_3$ or ${\cal D}_2$ would more
effectively reduce spurious reflections when octupolar contributions
are taken into account, we have calculated the reflection coefficient
due to backscatter of quadrupolar waves for ${\cal C}_3$:
\begin{equation}
\Big| \gamma\left( kR,\frac{2M}{R} \right) \Big|
 = \frac{45}{4(kR)^5}\frac{M}{R} 
   \left[\frac{ \left( 6 C_7(kR) - 7 C_8(kR) \right)^2 
 + \left( 6 S_7(kR) - 7 S_8(kR) \right)^2}{(kR)^2 + 9} \right]^{1/2}
 + O\left( \frac{2M}{R} \right)^2.
\label{Eq:ReflCoeffC3}
\end{equation}
Using the asymptotic expansions of $C_n(z)$ and $S_n(z)$ given in
Sect. \ref{Sect:AbsorbingBCa}, we find that Eq. (\ref{Eq:ReflCoeffC3})
decays as $(2M/R)(kR)^{-7}$ as $kR\to\infty$. Although not calculated,
the reflection coefficient due to backscatter of octupolar waves for
${\cal C}_3$ is expected to decay at least as fast or faster. For the
boundary condition ${\cal D}_2$, we find that when octupolar radiation
is considered, the reflection coefficient is (to leading order in
$2M/R$) the same as that given in Eq. (96) of \cite{lBoS06} and falls
off as $(k R)^{-6}$ for large $k R$. We conclude that indeed ${\cal
C}_3$ would be more effective than ${\cal D}_2$ unless the octupolar
contributions of the gravitational radiation are significantly
suppressed. Using the results in \cite{jBwP73}, the backscatter
corrections to first order in $2M/R$ can in principle be calculated to
give ${\cal D}_3$ and even ${\cal D}_L$.

When applied to full nonlinear formulations of Einstein's field
equations, the boundary conditions ${\cal C}_L$ and ${\cal D}_2$ on
$\Psi_0$ should be used along with constraint-preserving boundary
conditions and boundary conditions on the gauge degrees of freedom so
that the resulting initial-boundary value problem is well posed. A
possible strategy for specifying boundary conditions on the gauge
degrees of freedom is to insure that the outer boundary is a metric
sphere throughout the evolution. This should yield a natural split of
the spacetime manifold into a two-dimensional manifold $\tilde{\cal
M}$ and the two-sphere $S^2$. If the outer boundary is not a metric
sphere, our boundary conditions can still be applied; however, the
above split of the manifold first needs to be identified. Well posed
initial-boundary value formulations incorporating the boundary
conditions ${\cal C}_L$ and ${\cal D}_2$, and their numerical
implementation, will be studied in future work.

\begin{acknowledgments}

We thank J. Bardeen for his feedback on the results presented in this
article, for his calculations comparing our results with those in
Ref. \cite{jBwP73}, and for providing the boundary condition ${\cal
C}_L$ in Eq. (\ref{Eq:C_L}). We also thank L. Lehner, R. Matzner,
A. Nerozzi, O. Reula, O. Rinne, E. Schnetter and M. Tiglio for many
helpful discussions during the course of this work.  LTB was supported
by NSF grant PHY 0354842 and by NASA grant NNG 04GL37G to the
University of Texas at Austin. OCAS was partially supported by grant
CIC-UMSNH-4.20.
\end{acknowledgments}

\appendix
\section{On the stability of non-local boundary conditions}

In this appendix, we analyze the following initial-boundary value
problem on the quarter space $\Omega_R := (0,\infty) \times
(-\infty,R)$, $R > 0$:
\begin{eqnarray}
& \ddot{u}(t,x) - u''(t,x) + V(t,x) u(t,x) = F(t,x),
& \hbox{$(t,x) \in \Omega_R$},
\label{Eq:Evolution}\\
& \dot{u}(t,R) + u'(t,R) 
 + \frac{1}{R^2}\int\limits_0^\infty k_R(t,s) u(s,R) ds = G(t),
& \hbox{$t > 0$}
\label{Eq:Boundary}\\
&u(0,x) = f(x), \qquad
\dot{u}(0,x) = g(x),
& \hbox{$x < R$}.
\label{Eq:Initial}
\end{eqnarray}
Here, a dot and a prime denote differentiation with respect to $t$ and
$x$, respectively, $F$ is a given smooth source function on
$\Omega_R$, $G$ is a given smooth source function on $(0,\infty)$ and
$f$ and $g$ are smooth initial data on $(-\infty,R)$. Furthermore, we
assume that the potential $V$ in Eq. (\ref{Eq:Evolution}) and the
integral kernel $k_R(.,.): (0,\infty)^2 \to \Real$ in
Eq. (\ref{Eq:Boundary}) satisfy the following three conditions:
\begin{enumerate}
\item[(i)] $V: \Omega_R \to \Real$ is measurable and bounded in the
following sense: There exists a smooth and bounded function $W:
(-\infty,R) \to \Real$ and a strictly positive constant $p > 0$ such
that $p/R^2 \leq W(x)$ for all $x < R$, $V(t,x) \leq W(x)$ for all
$(t,x)\in\Omega_R$ and
\begin{displaymath}
D:=\sup\limits_{(t,x)\in\Omega_R} \frac{R^2}{\sqrt{p}}
    \left[ W(x) - V(t,x) \right] < \infty.
\end{displaymath}
(If $V$ is smooth, bounded, does not depend on $t$ and satisfies $V(x)
\geq p/R^2$ for all $x < R$ we can take $W(x) = V(x)$ and $D=0$.)
\item[(ii)] $k_R$ satisfies the causality condition $k_R(t,s) = 0$ if $s > t$,
\item[(iii)] There is a constant $C > 0$ which is independent of $R$
such that
\begin{eqnarray}
\int\limits_0^\infty | k_R(t,s) | dt \leq C R && \hbox{for all $s > 0$},
\nonumber\\
\int\limits_0^\infty | k_R(t,s) | ds \leq C R && \hbox{for all $t > 0$}.
\nonumber
\end{eqnarray}
\end{enumerate}
By a suitable re-definition of $u$ and $F$, if necessary, we can
assume that the boundary condition is homogeneous, that is, $G \equiv
0$. In this case, and under the assumptions (i),(ii) and (iii) above,
we now show that a sufficiently smooth solution of the IBVP
(\ref{Eq:Evolution},\ref{Eq:Boundary},\ref{Eq:Initial}) which vanishes
for $x$ sufficiently negative fulfills the energy estimate
\begin{equation}
{\cal E}(t) \leq e^{\mu t/R} \left[ {\cal E}(0) 
 + a R\int\limits_0^t\int\limits_{-\infty}^R |F(s,r)|^2 dr ds \right],
\label{Eq:EnergyEstimate}
\end{equation}
where the energy is defined by
\begin{displaymath}
{\cal E}(t) = \frac{1}{2} \int\limits_{-\infty}^R 
\left( |\dot{u}(t,x) |^2 + |u'(t,x)|^2 + W(x) |u(t,x)|^2 \right) dx,
\end{displaymath}
and where the dimensionless constants $a$ and $\mu$ are given by
\begin{displaymath}
a = \frac{1}{2\delta}\, ,\qquad
\mu = \delta + D + \frac{C^2}{2\sqrt{p}}\, ,
\end{displaymath}
$\delta$ being a strictly positive but otherwise arbitrary
constant. The estimate (\ref{Eq:EnergyEstimate}) proves uniqueness and
continuous dependence of the solution on the data. We do not prove
existence of solutions here.

In order to prove the estimate (\ref{Eq:EnergyEstimate}) we first
differentiate ${\cal E}(t)$ with respect to $t$ and use the evolution
equation (\ref{Eq:Evolution}) and assumption (i) and obtain
\begin{eqnarray}
\frac{d}{dt} {\cal E}(t) &=& \int\limits_{-\infty}^R \left\{
 \dot{u}(t,x) u''(t,x) + \dot{u}'(t,x) u'(t,x) 
 + \left[ W(x) - V(t,x) \right]\dot{u}(t,x) u(t,x) 
 + \dot{u}(t,x) F(t,x) \right\} dx
\nonumber\\
 &\leq& \dot{u}(t,R) u'(t,R) 
 + \frac{D + \delta}{R} {\cal E}(t) 
 + \frac{R}{2\delta}\int\limits_{-\infty}^R |F(t,x)|^2 dx,
\nonumber
\end{eqnarray}
where $\delta > 0$ and where we have used Schwarz's inequality in the
last step. Integrating over $t$ and using the boundary condition
(\ref{Eq:Boundary}), we obtain
\begin{eqnarray}
{\cal E}(t) - {\cal E}(0) &\leq& -\int\limits_0^t |u'(s,R)|^2 ds
  - \frac{1}{R^2}\int\limits_0^t \int\limits_0^\infty 
    k_R(s,\tau) u'(s,R) u(\tau,R) d\tau ds
\nonumber\\
 &+& \frac{D + \delta}{R}\int\limits_0^t {\cal E}(s) ds
 + \frac{R}{2\delta}\int\limits_0^t \int\limits_{-\infty}^R |F(s,x)|^2 dx ds.
\label{Eq:Est1}
\end{eqnarray}
In order to estimate the second term on the right-hand side, we use
assumption (iii) and Schwarz's inequality again, and obtain
\begin{eqnarray}
\left| \int\limits_0^t \int\limits_0^\infty
           k_R(s,\tau) u'(s,R) u(\tau,R) d\tau ds \right|
 &\leq& \int\limits_0^t |u'(s,R)| \left( 
 \int\limits_0^\infty |k_R(s,\tau)|^{1/2} |k_R(s,\tau)|^{1/2} |u(\tau,R)| d\tau
 \right) ds
\nonumber\\
 &\leq& \int\limits_0^t |u'(s,R)| 
 \left( \int\limits_0^\infty |k_R(s,\tau)| d\tau \right)^{1/2}
 \left( \int\limits_0^\infty |k_R(s,\tau)| |u(\tau,R)|^2 d\tau \right)^{1/2}
 ds
\nonumber\\
 &\leq& (C R)^{1/2} \left( \int\limits_0^t |u'(s,R)|^2 ds \right)^{1/2}
  \left( \int\limits_0^t \int\limits_0^\infty 
              |k_R(s,\tau)| |u(\tau,R)|^2 d\tau ds \right)^{1/2}.
\nonumber
\end{eqnarray}
Using Fubini's theorem, the causality condition (ii) and condition
(iii), we have
\begin{displaymath}
\int\limits_0^t \int\limits_0^\infty |k_R(s,\tau)| |u(\tau,R)|^2 d\tau ds
 = \int\limits_0^\infty \int\limits_0^t |k_R(s,\tau)| ds |u(\tau,R)|^2 d\tau
 = \int\limits_0^t \int\limits_0^t |k_R(s,\tau)| ds |u(\tau,R)|^2 d\tau
 \leq C R \int\limits_0^t |u(\tau,R)|^2 d\tau.
\end{displaymath}
Therefore,
\begin{equation}
\left| \int\limits_0^t \int\limits_0^\infty
           k_R(s,\tau) u'(s,R) u(\tau,R) d\tau ds \right|
 \leq C R \left( \int\limits_0^t |u'(s,R)|^2 ds \right)^{1/2}
          \left( \int\limits_0^t |u(s,R)|^2 ds \right)^{1/2}.
\label{Eq:Est2}
\end{equation}
The estimates (\ref{Eq:Est1},\ref{Eq:Est2}) imply that
\begin{eqnarray}
{\cal E}(t) - {\cal E}(0) &\leq&
 - \left[ \left( \int\limits_0^t |u'(s,R)|^2 ds \right)^{1/2}
    - \frac{C}{2R}\left( \int\limits_0^t |u(s,R)|^2 ds \right)^{1/2} \right]^2
\nonumber\\
&+& \frac{C^2}{4R^2} \int\limits_0^t |u(s,R)|^2 ds
 +  \frac{D + \delta}{R}\int\limits_0^t {\cal E}(s) ds
 + \frac{R}{2\delta}\int\limits_0^t \int\limits_{-\infty}^R |F(s,x)|^2 dx ds
\nonumber\\
 &\leq& \frac{C^2}{4R^2} \int\limits_0^t |u(s,R)|^2 ds
 +  \frac{D + \delta}{R}\int\limits_0^t {\cal E}(s) ds
 + \frac{R}{2\delta}\int\limits_0^t \int\limits_{-\infty}^R |F(s,x)|^2 dx ds.
\nonumber
\end{eqnarray}
Finally, using the Sobolev type estimate
\begin{displaymath}
|u(s,R)|^2 = \int\limits_{-\infty}^R \partial_x [u(s,x)^2] dx
           = 2\int\limits_{-\infty}^R u(s,x) u'(s,x) dx
        \leq \frac{2R}{\sqrt{p}}\, {\cal E}(s),
\end{displaymath}
we obtain the estimate (\ref{Eq:EnergyEstimate}) after the application
of Gronwall's lemma.

As an application of our result, consider the massless scalar wave
equation on a Schwarzschild background of mass $M$. Performing a
decomposition of the scalar field $\Phi$ into spherical harmonics
$Y^{\ell m}$,
\begin{displaymath}
\Phi(t,r,\vartheta,\varphi) = \frac{1}{r}
\sum\limits_{\ell=0}^\infty\sum\limits_{m=-\ell}^\ell
\phi_{\ell m}(t,r) Y^{\ell m}(\vartheta,\varphi),
\end{displaymath}
the wave equation reads
\begin{equation}
\left[ \partial_t^2 - \partial_{r_*}^2 + \left( 1 - \frac{2M}{r} \right)
\left( \frac{\ell(\ell+1)}{r^2} + \frac{2M}{r^3} \right) \right] 
\phi_{\ell m} = 0,
\label{Eq:MassLessWave}
\end{equation}
In outgoing Eddington-Finkelstein coordinates $\tau = t + r - r_*$,
$\rho = r$, this equation assumes the form
\begin{displaymath}
\left[ \partial_\tau^2 - \partial_\rho^2 + \frac{\ell(\ell+1)}{\rho^2} \right]
\phi_{\ell m} = -\frac{2M}{\rho}\left[ (\partial_\tau + \partial_\rho)^2
 - \frac{1}{\rho}(\partial_\tau + \partial_\rho) + \frac{1}{\rho^2}
\right] \phi_{\ell m}\, .
\end{displaymath}
Monopolar outgoing solutions have the form (omitting the indices $00$
on $\phi$)
\begin{displaymath}
\phi(t,r) = U(r_* - t) + 2M\int\limits_{r_* - t}^\infty 
 \frac{U(x) dx}{(t + 2r - r_* + x)^2} 
 + O\left( \frac{2M}{R} \right)^2,
\end{displaymath}
where $U$ is a smooth and bounded function which vanishes if its
argument is sufficiently negative. Since
\begin{displaymath}
(\partial_t + \partial_{r_*})\phi(t,r) = -\frac{2M}{r^2}
 \int\limits_0^\infty \frac{U(r_* - t + 2ry)dy}{(1+y)^3}
 + O\left( \frac{2M}{R} \right)^2,
\end{displaymath}
the boundary condition 
\begin{equation}
(\partial_t + \partial_{r_*})\phi(t,R)  
 + \frac{2M}{R^2}
   \int\limits_0^{t/2R} \frac{\phi(t - 2R y,R) dy}{(1+y)^3} = G(t),
\label{Eq:BoundaryBackModel}
\end{equation}
where
\begin{displaymath}
G(t) = -\frac{2M}{R^2}\int\limits_{t/2R}^\infty 
   \frac{\phi(0,(R_* + 2Ry - t)_\bullet,R) dy}{(1+y)^3}\; ,
\end{displaymath}
is perfectly absorbing up to order $2M/R$. Performing the
substitutions $x = r_*$ and $y = (t-s)/2R$ equation
(\ref{Eq:MassLessWave}) can be brought into the form
(\ref{Eq:Evolution}) and the boundary condition
(\ref{Eq:BoundaryBackModel}) into the form (\ref{Eq:Boundary}), where
\begin{displaymath}
k_R(t,s) = \left\{ \begin{array}{ll} 
 \frac{M}{R}\frac{(2R)^3}{(2R + t - s)^3}\, , & 0 \leq s \leq t, \\
                                        0   , & s > t.
\end{array} \right.
\end{displaymath}
This kernel satisfies the assumptions (ii) and (iii) with $C = M/R$,
and the potential in Eq. (\ref{Eq:MassLessWave}) is smooth, positive
and bounded by the constant $W:=[\ell(\ell+1)+1]/(2M)^2$; hence, it
satisfies assumption (i). For linear fluctuations of a Schwarzschild
background, Eq. (\ref{Eq:MassLessWave}) has to be replaced by the
Regge-Wheeler or Zerilli equation (\ref{Eq:RWZ}), and our boundary
condition ${\cal D}_2$ involves third derivatives of the
fields. Therefore, it is not of the form (\ref{Eq:Boundary}). However,
it can be reduced to a first order system for which an energy estimate
similar to the one presented above can be found by introducing
auxiliary variables as in the method used in the appendix of
Ref. \cite{lBoS06}.

\bibliography{refs}

\begin{thebibliography}{10}

\bibitem{hFgN99}
H.~Friedrich and G.~Nagy.
\newblock The initial boundary value problem for {E}instein's vacuum field
  equations.
\newblock {\em Comm. Math. Phys.}, 201:619--655, 1999.

\bibitem{lKlLmSlBhP05}
L.E. Kidder, L.~Lindblom, M.A. Scheel, L.T. Buchman, and H.P. Pfeiffer.
\newblock Boundary conditions for the {E}instein evolution system.
\newblock {\em Phys. Rev. D}, 71:064020(1)--064020(22), 2005.

\bibitem{oSmT05}
O.~Sarbach and M.~Tiglio.
\newblock Boundary conditions for {E}instein's field equations: Mathematical
  and numerical analysis.
\newblock {\em Journal of Hyperbolic Differential Equations}, 2:839--883, 2005.

\bibitem{lLmSlKrOoR06}
L.~Lindblom, M.A. Scheel, L.E. Kidder, R.~Owen, and O.~Rinne.
\newblock A new generalized harmonic evolution system.
\newblock {\em Class. Quantum Grav.}, 23:S447--S462, 2006.

\bibitem{oR06}
O.~Rinne.
\newblock Stable radiation-controlling boundary conditions for the generalized
  harmonic {E}instein equations.
\newblock {\em Class. Quantum Grav.}, 23:6275--6300, 2006.

\bibitem{gNoS06}
G.~Nagy and O.~Sarbach.
\newblock A minimization problem for the lapse and the initial-boundary value
  problem for {E}instein's field equations.
\newblock {\em Class. Quantum Grav.}, 23:S477--S504, 2006.

\bibitem{mBbSjW06}
M.C. Babiuc, B.~Szilagyi, and J.~Winicour.
\newblock Harmonic initial-boundary evolution in general relativity.
\newblock {\em Phys. Rev. D}, 73:064017(1)--064017(23), 2006.

\bibitem{hKjW06}
H.O. Kreiss and J.~Winicour.
\newblock Problems which are well-posed in a generalized sense with
  applications to the {E}instein equations.
\newblock {\em Class. Quantum Grav.}, 23:S405--S420, 2006.

\bibitem{mBhKjW07}
M.C. Babiuc, H-O. Kreiss, and J.~Winicour.
\newblock Constraint-preserving {S}ommerfeld conditions for the harmonic
  {E}instein equations.
\newblock {\em Phys. Rev. D}, 75:044002(1)--044002(13), 2007.

\bibitem{bZePpDmT06}
B.~Zink, E.~Pazos, P.~Diener, and M.~Tiglio.
\newblock Cauchy-perturbative matching revisited: Tests in spherical symmetry.
\newblock {\em Phys. Rev. D}, 73:084011(1)--084011(14), 2006.

\bibitem{jNsB04}
J.~Novak and S.~Bonazzola.
\newblock Absorbing boundary conditions for simulation of gravitational waves
  with spectral methods in spherical coordinates.
\newblock {\em J. Comp. Phys.}, 197:186--196, 2004.

\bibitem{sBeGpG04}
S.~Bonazzola, E.~Gourgoulhon, P.~Grandcl\a'ement, and J.~Novak.
\newblock Constrained scheme for the {E}instein equations based on the {D}irac
  gauge and spherical coordinates.
\newblock {\em Phys. Rev. D}, 70:104007(1)--104007(24), 2004.

\bibitem{mClLiOrPfPhV03}
M.~Choptuik, L.~Lehner, I.~Olabarrieta, R.~Petryk, F.~Pretorius, and
  H.~Villegas.
\newblock Towards the final fate of an unstable black string.
\newblock {\em Phys. Rev. D}, 68:044001(1)--044001(11), 2003.

\bibitem{fP05}
F.~Pretorius.
\newblock Evolution of binary black-hole spacetimes.
\newblock {\em Phys. Rev. Lett.}, 95:121101(1)--121101(4), 2005.

\bibitem{hF81}
H.~Friedrich.
\newblock On the regular and asymptotic characteristic initial value problem
  for {E}instein's vacuum field equations.
\newblock {\em Proc. R. Soc. Lond. A}, 375:169--184, 1981.

\bibitem{jF98}
J.~Frauendiener.
\newblock Numerical treatment of the hyperboloidal initial value problem for
  the vacuum {E}instein equations. 2. the evolution equations.
\newblock {\em Phys. Rev. D}, 58:064003(1)--064003(18), 1998.

\bibitem{sHcStVaZ05}
S.~Husa, C.~Schneemann, T.~Vogel, and A.~Zengino\u{g}lu.
\newblock Hyperboloidal data and evolution.
\newblock {\em AIP Conf. Proc.}, 841:306--313, 2006.

\bibitem{lBoS06}
L.T. Buchman and O.C.A. Sarbach.
\newblock Towards absorbing outer boundaries in general relativity.
\newblock {\em Class. Quantum Grav.}, 23:6709--6744, 2006.

\bibitem{sL04a}
S.R. Lau.
\newblock Rapid evaluation of radiation boundary kernels for time-domain wave
  propagation on blackholes: theory and numerical methods.
\newblock {\em J. Comput. Phys.}, 199:376--422, 2004.

\bibitem{sL04b}
S.R. Lau.
\newblock Rapid evaluation of radiation boundary kernels for time-domain wave
  propagation on black holes: implementation and numerical tests.
\newblock {\em Class. Quantum Grav.}, 21:4147--4192, 2004.

\bibitem{sL05}
S.R. Lau.
\newblock Analytic structure of radiation boundary kernels for blackhole
  perturbations.
\newblock {\em J. Math. Phys.}, 46:102503(1)--102503(21), 2005.

\bibitem{ePeDaNcPeSmT07}
E.~Pazos, E.N. Dorband, A.~Nagar, C.~Palenzuela, E.~Schnetter, and M.~Tiglio.
\newblock How far away is far enough for extracting numerical waveforms, and
  how much do they depend on the extraction method?, 2006.
\newblock gr-qc/0612149.

\bibitem{oSmT01}
O.~Sarbach and M.~Tiglio.
\newblock Gauge invariant perturbations of {S}chwarzschild black holes in
  horizon penetrating coordinates.
\newblock {\em Phys. Rev. D}, 64:084016(1)--(15), 2001.

\bibitem{jB07a}
J.M. Bardeen, 2007.
\newblock Private communication.

\bibitem{jBwP73}
J.M. Bardeen and W.H. Press.
\newblock Radiation fields in the {S}chwarzschild background.
\newblock {\em J. Math. Phys.}, 14:7--19, 1973.

\bibitem{tRjW57}
T.~Regge and J.~Wheeler.
\newblock Stability of a {S}chwarzschild singularity.
\newblock {\em Phys. Rev.}, 108:1063--1069, 1957.

\bibitem{fZ70}
F.~Zerilli.
\newblock Effective potential for even-parity {R}egge-{W}heeler gravitational
  perturbation equations.
\newblock {\em Phys. Rev. Lett.}, 24:737--738, 1970.

\bibitem{Teukolsky73}
S.A. Teukolsky.
\newblock Perturbations of a rotating black hole. 1. fundamental equations for
  gravitational electromagnetic, and neutrino field perturbations.
\newblock {\em Astrophys. J.}, 185:635--647, 1973.

\bibitem{uGuS79}
U.H. Gerlach and U.K. Sengupta.
\newblock Gauge-invariant perturbations on most general spherically symmetric
  space-times.
\newblock {\em Phys. Rev. D}, 19:2268--2272, 1979.

\bibitem{rP72b}
R.H. Price.
\newblock Nonspherical perturbations of relativistic gravitational collapse.
  {II. I}nteger-spin, zero-rest-mass fields.
\newblock {\em Phys. Rev. D}, 5:2439--2454, 1972.

\bibitem{rP65}
R.~Penrose.
\newblock Zero rest-mass fields including gravitation: asymptotic behaviour.
\newblock {\em Proc. R. Soc. Lond. A}, 284:159--203, 1965.

\bibitem{Chandrasekhar}
S.~Chandrasekhar.
\newblock {\em The Mathematical Theory of Black Holes}.
\newblock Oxford University Press, Great Clarendon Street, Oxford 0X2 6DP,
  1992.

\end{thebibliography}
\end{document}